\documentclass[a4paper, 11pt]{article}
\usepackage{amsmath, amssymb, graphics}
\usepackage{longtable}
\usepackage{lscape}
\newcommand{\mathsym}[1]{{}}
\newcommand{\unicode}[1]{{}}

\catcode`\@=11

\@addtoreset{equation}{section} 

\def\d{\mbox{\rm d}}

\def\dddot#1{\mathinner{\buildrel\vbox{\kern5pt\hbox{...}}\over{#1}}}
\def\ddddot#1{\mathinner{\buildrel\vbox{\kern5pt\hbox{....}}\over{#1}}}

\begin{document}

\begin {center}
{\Large Lie Symmetries and Similarity transformations for the Generalized Boiti-Leon-Pempinelli equations}\\[3 mm]
{\small K. Krishnakumar$^{1}$, A. Durga Devi$^{2}$ and A. Paliathanasis}$%
^{3,4}${\small \\[3mm]
$^{1}$ Department of Mathematics, Srinivasa Ramanujan Centre, SASTRA Deemed
to be University, Kumbakonam 612 001, India.\\[0pt]
$^{2}$ Department of Physics, Srinivasa Ramanujan Centre, SASTRA Deemed to be University, Kumbakonam 612 001, India.}\\[0pt]
$^{3}${\small Institute of Systems Science, Durban University of Technology,
Durban 4000, South Africa}\\[0pt]
$^{4}${\small Instituto de Ciencias F\'{\i}sicas y Matem\'{a}ticas,
Universidad Austral de Chile, Valdivia 5090000, Chile}
\end{center}

\begin{center}
\textbf{Abstract:} We perform a detailed classification of the Lie point symmetries and of the resulting similarity transformations for the Generalized Boiti-Leon-Pempinelli equations. The latter equations for a system of two nonlinear 1+2 partial differential equations of second- and third-order. The nonlinear equations depend of two parameters, namely $n$ and $m$, from where we find that for various values of these two parameters the resulting systems admit different number of Lie point symmetries. For every case, we present the complete analysis for the admitted Lie group as also we determine all the possible similarity solutions which follow from the one-dimensional optimal system. Finally, we summarize the results by presenting them in a tabular way.
\end{center}


\strut\hfill

\section{Introduction}

Nonlinear ordinary and partial differential equations are playing very
prominent role in field of Mathematics, Physics, Mathematical Physics and
Mathematical modeling etc.. Through the study of solutions of such
differential equations one can understand the behavior or consequence of a
natural phenomena very clearly. Though various methods are available in the
literature, Lie' symmetry method successfully using by researchers due to
its systematic and algorithmic way to derive the solution of the
differential equations. Actually this method was developed by Sophus Lie,
during the period $1872-1899$ \cite{Lie 70 a, Lie 70 b, Lie 70 c, Lie 67 a,
Lie 71 a, Lie 77 a, Lie 12 a} and applied it successfully without any ansatz
to solve the differential equations.

Since 1960 over the explicit construction of solutions of any sort of
problems, even complicated, of mathematical physics, this theory was
exploited by the Russian school with L. V. Ovsiannikov \cite{Ovsyannikov 58
a, Ovsyannikov 59 a, Ovsiannikov 82 a}. During the last few decades, Lie's
theory continuously enhanced the researchers to solve the differential
equations in various form such as generalized symmetries \cite{Ovsiannikov 82 a, Olver 93 a}, approximate symmetries \cite{Baikov 89,
Ibraghimov 09},
and nonlocal symmetries \cite{Olver 93 a, Bluman 89 a, Olver 03 a, Leach 07}
to quote a few. Therefore, researchers using the method widely on both
theoretical and applied point of view \cite{Bluman 69 a, Bluman 69 b,
BluC74, Bluman 88 a, Bluman 89 a, Ibragimov 83 a, Ibragimov 99 a, nailio,
Ibraghimov, Olver 77 a, Olver 86 a, Olver 93 a, Olver 02 a,Olver 03 a,
Andriopoulos 06 a, Leach 78 a, Leach 80 e, Leach 80 f, Leach 81 a, Leach 03
c}. Nowadays researchers are using powerful Computer Algebra Systems
(CAS) like Maple and Mathematica (commercial), etc. to do the calculations
over the symmetry rapidly. In this work, for the calculation of the
symmetries we use the Mathematica add-on Sym \cite{Dimas 05 a, Dimas 06 a,
Dimas 08 a, Andriopoulos 09 a}.

In this work we are interested on the complete classification of the Lie
symmetries and of the one-dimensional optimal system for the nonlinear
system of 1+2 partial differential equations known as generalized
Boiti-Leon-Pempinelli (BLP) equations, the original equations are \cite{blp1}%
\begin{eqnarray*}
u_{ty}-\left( u^{2}-u_{x}\right) _{xy}-2v_{xxx} &=&0 ,\\
v_{t}-v_{xx}-2uv_{x} &=&0.
\end{eqnarray*}%
while in this work we are interesting on the generalization
\begin{eqnarray}
u_{ty}-\left( u^{n}-u_{x}\right) _{xy}-\alpha v_{xxx} &=&0,  \label{1} \\
v_{t}-v_{xx}-\beta u^{m}v_{x} &=&0,  \notag
\end{eqnarray}%
where $nm\neq 0$ and $\alpha \beta \neq 0.$

The BLP equation describes interactions of two waves with different
dispersion relations and are the two-dimensional generalizations of the
sine- and sinh-Gordon equations. Some exact and analytic solutions
determined in \cite{blp2,blp3,blp4,blp5a,blp5b}, while recently in \cite{Zhao17}
a complete analysis of the Lie point symmetries and of the B\"{a}cklund
transformations performed.

In this work we classify the admitted Lie point symmetries of the
generalized BLP equations for various values of the free parameters $n,m$,
while the latter are constrained by the theory of Lie symmetries. Such
analysis is important for the detailed study of nonlinear partial
differential equations and for the determination of new exact solutions, for
other examples we refer the reader in \cite{app1,app2,app3,app4} and
references therein. The plan of the paper is as follows.

In\ Section 2, we present the basic properties and definitions for the
theory of Lie point symmetries of differential equations. We discuss the
main mathematical methods which are applied in this work. Section 3 includes
the main analysis of this work where we present the complete classification
of the Lie point symmetries for system (\ref{1}). We found four different
possible cases for the set of variables $\left\{ n,m\right\} $. For every
set we determine the Lie point symmetries and we present the commutators and
the adjoint representation. The latter are used for the derivation of the
one-dimensional optimal system which are applied for the derivation of new
similarity solutions for the generalized \ BLP equations (\ref{1}).
Finally, our results are summarized in Section 4, where we draw our
conclusions.

\section{Lie's Theory}

Consider a system of equation as follows
\begin{eqnarray}
\triangle
_{1}(t,x,y,u,v,u_{t},v_{t},u_{x},v_{x},u_{y},v_{y},u_{tt},v_{tt},u_{tx},v_{tx},u_{ty},v_{ty},u_{xx},u_{xy},...)=0, &&
\notag \\
&&  \label{G1} \\
\triangle
_{2}(t,x,y,u,v,u_{t},v_{t},u_{x},v_{x},u_{y},v_{y},u_{tt},v_{tt},u_{tx},v_{tx},u_{ty},v_{ty},u_{xx},u_{xy},...)=0, &&
\notag \\
&&  \label{G2}
\end{eqnarray}%
where $t,x,y$ are taken to represent the independent variables and $u,v$ are
taken to represent dependent variables, that is $u=u\left( t,x,y\right) $
and $v=v\left( t,x,y\right) $, while $u_{t}=\frac{\partial u}{\partial t}$.

Infinitesimal point transformation for each variables is defined as in the
following manner,
\begin{equation*}
\tilde{t}(t,x,\ y,\ \epsilon )=t+\epsilon \xi ^{1}(t,x,\ y)+\circ (\epsilon
^{2})=t+\epsilon \xi ^{1}\left( t,x,y\right) +\circ (\epsilon ^{2}),
\end{equation*}%
\begin{equation*}
\tilde{x}(t,x,\ y,\ \epsilon )=x+\epsilon \xi ^{2}(t,x,\ y)+\circ (\epsilon
^{2})=x+\epsilon \xi ^{2}\left( t,x,y\right) +\circ (\epsilon ^{2}),
\end{equation*}%
\begin{equation*}
\tilde{y}(t,x,\ y,\ \epsilon )=y+\epsilon \xi ^{3}(t,x,\ y)+\circ (\epsilon
^{2})=y+\epsilon \xi ^{3}\left( t,x,y\right) +\circ (\epsilon ^{2}),
\end{equation*}%
\begin{equation*}
\tilde{u}(t,x,\ y,\ \epsilon )=u+\epsilon \eta ^{1}(t,x,\ y)+\circ (\epsilon
^{2})=u+\epsilon \eta ^{1}\left( t,x,y\right) +\circ (\epsilon ^{2}),
\end{equation*}%
\begin{equation*}
\tilde{v}(t,x,\ y,\ \epsilon )=v+\epsilon \eta ^{2}(t,x,\ y)+\circ (\epsilon
^{2})=v+\epsilon \eta ^{2}\left( t,x,y\right) +\circ (\epsilon ^{2}),
\end{equation*}%
where $X$ is called infinitesimal generator defined as
\begin{equation*}
X=\frac{\partial \tilde{t}}{\partial \epsilon }\partial _{t}+\frac{\partial
\tilde{x}}{\partial \epsilon }\partial _{x}+\frac{\partial \tilde{y}}{%
\partial \epsilon }\partial _{y}+\frac{\partial \tilde{u}}{\partial \epsilon
}\partial _{u}+\frac{\partial \tilde{v}}{\partial \epsilon }\partial _{v},
\end{equation*}%
or equivalently
\begin{equation*}
X=\xi ^{1}(t,x,\ y)\partial _{t}+\xi ^{2}(t,x,\ y)\partial _{x}+\xi
^{3}(t,x,\ y)\partial _{y}+\eta ^{1}(t,x,\ y)\partial _{u}+\eta ^{2}(t,x,\
y)\partial _{v}.
\end{equation*}

The invariant condition for the system (\ref{G1}) and (\ref{G2}), based on
the Lie's theory is given by
\begin{eqnarray*}
\triangle _{1}(t,x,y,u,v)=\triangle _{1}(\tilde{t},\tilde{x},\tilde{y},%
\tilde{u},\tilde{v}) && \\
\triangle _{2}(t,x,y,u,v)=\triangle _{2}(\tilde{t},\tilde{x},\tilde{y},%
\tilde{u},\tilde{v}) &&
\end{eqnarray*}%
By using infinitesimal transformation and invariant condition one can find
the infinitesimal generator which is known as symmetries of the system (\ref%
{G1}) and (\ref{G2}). Therefore, if $X$ is a Lie point symmetry for equation
$\triangle _{1}\equiv 0$ and $\triangle _{2}\equiv 0$ then the following
condition is true%
\begin{eqnarray*}
X^{\left[ k\right] }\triangle _{1}=\lambda \triangle _{1}~,~\text{mod}%
\triangle _{1} &=&0 \\
X^{\left[ k\right] }\triangle _{2}=\phi \triangle _{2}~,~\text{mod}\triangle
_{2} &=&0
\end{eqnarray*}%
where $\lambda $ and $\phi $ are arbitrary functions, and $X^{\left[ k\right]
}$ is the k-th extension of $X$ in the jet-space. These symmetries play a
crucial role to perform the reduction of the order of the differential
equations as well as number of independent variables. The latter system
known as determining system provides the Lie point symmetries for a given
system of differential equations.

\subsection{Lie invariants}
By applying the  Lie's theory one can find the symmetries of a given differential equations. By using the Lie symmetry with some additional regularity assumptions we obtain transformations which is called Lie's invariant. The invariant reduces the given equation into another which involve fewer independent variables than the original equation. The solution of the reduced equation is called invariant solution to the given equation.

\subsection{One-dimensional optimal system}
Optimal system forms by a list of $n-$parameter subalgebras if every $n-$parameter subalgebras is equivalent to a unique member of the list under some element of the adjoint representation. P.J. Olver discussed the procedure to find adjoint representation and optimal system which are giving an unique combination of symmetries to perform the reductions of a given differential equation \cite{Olver 86 a}. Later researcher follows slightly different method from Olver which proposed by Ibragimov  \cite{Ibragimov 08 c}. The method is given a simple way to find the optimal system \cite{Grigoriev 10, Liu 10, Zhao15, Zhao17}.

\section{Lie symmetry classification}

We apply the theory of Lie point symmetries and we have obtained four sets
for the set of variables $\left\{ n,m\right\} $ where we find the symmetries
of the system ({\ref{1}}). In particular we found that of $\left\{
n,m\right\} =\left( 2,1\right) $ which is the original 1+2 BLP equation, for
$\left\{ n,m\right\} =\left\{ 1,1\right\} $ and for $\left\{ n,m\right\}
=\left( n,n-1;n\neq 1\right) $ the admitted Lie point symmetries are three
plus infinity plus infinity but with different admitted Lie algebra. On the
other hand for arbitrary $\left\{ n,m\right\} $ the admitted Lie point
symmetries are the common elements, of the common subalgebra for the
previous cases.

For each case we present the commutator table and the adjoint
representation. From the latter results we determine the one-dimensional
optimal system which is applied for the derivation of all the possible
similarity transformations and reduction, and when it is feasible of exact
solutions.

\subsection{Case $n=2,$ $m=1$}

If $m=1$ and $n=2$ then the system (\ref{1}) yields the following equation
\begin{eqnarray}\label{1.12}
u_{ty}-\left( u^{2}-u_{x}\right) _{xy}-\alpha v_{xxx} &=&0, \nonumber\\
v_{t}-v_{xx}-\beta u v_{x} &=&0.
\end{eqnarray}

The admitted Lie symmetries are%
\begin{eqnarray*}
X_{1} &=&\partial _{t}~,~X_{2}=\partial _{x}~,~X_{3}=\psi(y)\partial
_{y}-\psi^\prime v \partial_v~,~X_{4}^{\infty }=\phi \left( y\right) \partial _{v}~, \\
X_{5} &=&2t \partial _{t}+x \partial _{x}-u \partial _{u}.
\end{eqnarray*}
The commutator table and adjoint representation based on the formulas $[X_m,X_n]=X_mX_n-X_nX_m$ and $Ad \left[ e^{\epsilon\textbf{X}_{I}}\right]\textbf{X}_{J}=X_J-\epsilon[X_I,X_J]
	+\frac{\epsilon^2}{2}[X_I,[X_I,X_J]]-... $ respectively are given by the following tables $(\ref{T1})$ and $(\ref{T2})$.

\begin{table}[htb]\label{T1}
\caption{Commutator Table for case 1.1 with $\psi=1$ }\label{T1}
\begin{center}
\begin{tabular}{c||ccccc}
\hline\hline
$\left[ X_{I},X_{J}\right] $ & $X_{1}$ & $X_{2}$ & $X_{3}$ & $X_{4}$ & $X_{5}$ \\
\hline
$X_{1}$ & $0$ & $0$ & $0$ & $0$& $2X_{1}$  \\
$X_{2}$ & $0$ & $0$ & $0$  & $0$& $X_{2}$\\
$X_{3}$ & $0$ & $0$  & $0$& $(\psi^{\prime}\phi+\phi^{\prime}\psi)\partial_v$ & $0$\\
$X_{4}$ & $0$ & $0$  & $-(\psi^{\prime}\phi+\phi^{\prime}\psi)\partial_v$& $0$ & $0$\\
$X_{5}$ & $-2X_{1}$ & $-X_{2}$  & $0$& $0$ & $0$\\
\hline\hline
\end{tabular}
\end{center}
\end{table}

\begin{table}[htb]
\caption{Adjoint representation Table for case 1.1 with $\psi=1$ }\label{tableExample}\label{T2}
\begin{center}
\begin{tabular}{c||cccc}
\hline\hline
$\left[ Ad(e^{\epsilon X_i})X_j\right] $ & $X_{1}$ & $X_{2}$ & $X_{3}$ & $X_{5}$ \\
\hline
$X_{1}$ & $X_{1}$ & $X_{2}$ & $X_{3}$ & $X_{5}-2\epsilon X_{1}$  \\
$X_{2}$ & $X_{1}$ & $X_{2}$ & $X_{3}$  & $X_{5}-\epsilon X_{2}$\\
$X_{3}$ & $X_{1}$ & $X_{2}$  & $X_{3}$ & $X_{5}$\\
$X_{5}$ & $e^{2\epsilon}X_{1}$ & $e^{\epsilon}X_{2}$  & $X_{3}$& $X_{5}$\\
\hline\hline
\end{tabular}
\end{center}
\end{table}

\begin{landscape}
\begin{table}[htb]
\caption{Reductions for the case $m=1,n=2$ }\label{tableExample}\label{T3}
\begin{center}
\begin{tabular}{c|c|cc}
\hline\hline
Optimal System & Similarity variable & &Reductions  \\
\hline
$ X_5+c_1X_3$ & $X=\dfrac{x}{\sqrt{t}},Y=y-\frac{c_1}{2}\log t,$ & $(I)$& $\alpha V_{XXX}-U_{XXY}+2U U_{XY}+2U_X U_Y+$  \\

 & $u=\dfrac{1}{\sqrt{t}}U\left[X,Y\right],
v=V\left[X,Y\right]$ & & $\dfrac{1}{2}X U_{XY}+\frac{c_1}{2}U_{YY}+\frac{1}{2}U_Y=0$\\

 &  &  &$V_{XX}+\beta U V_X+\frac{1}{2}X V_X+\frac{c_1}{2} V_Y=0$ \\
\hline

$X_1+c_1 X_2 +c_2 X_3$ & $r=t+c_1x+c_2y, u=U(r), v=V(r)$ & $(II)$&$c_2 U^{\prime\prime}-c_1c_2\left(2{U^{\prime}}^2+2UU^{\prime\prime}\right)
+{c_1}^2c_2U^{\prime\prime\prime}+\alpha{c_1}^3V^{\prime\prime\prime}=0
$ \\

 &  & &$(1-c_1\beta U)V^{\prime}-{c_1}^2V^{\prime\prime}=0$ \\
\hline
$X_{1}$ & $u=U[x,y],v=V[x,y]$ & $(III)$&$2(U_x U_y+UU_{xy})-U_{xxy}+\alpha V_{xxx}=0$
 \\

& & &$\beta U V_x+V_{xx}=0$   \\
\hline
$X_{2}$ & $u(t,x,y)=U(t,y), v(t,x,y)=V(t,y)$ & $(IV)$&$U_{ty}=0, $ \\
& & &$V_t=0 $  \\ \hline
$X_{3}$ & $u(t,x,y)=U(t,x),v(t,x,y)=V(t,x).$ & $(V)$& $\beta U V_x+V_{xx}-V_t=0,
$  \\
 & & &$V_{xxx}=0$  \\ \hline
$X_{5}$ & $X=x\sqrt t, Y=y,$ & $(VI)$&$2\alpha V_{XXX}+ (1+4U_X)U_Y+(X+4U) U_{XY}-2U_{YXX}=0$  \\
 & $u(t,x,y)=\frac{1}{\sqrt t}U(X,Y), v(t,x,y)=V(X,Y)$  & &$(X+2\beta U)V_X+2V_{XX}=0$  \\ \hline\hline
\end{tabular}
\end{center}
\end{table}
\end{landscape}

As discussed in \cite{Liu 10} the optimal system are $ X_5+c_1X_3,~ X_1+c_1 X_2 +c_2 X_3,     X_1, X_2 , X_3, X_5$. Now we have investigated invariant solution for (\ref{1.12}) corresponding to each optimal system and tabulated them in table (\ref{T3}). For example the characteristic equation of $X_5+c_1X_3$ is written as
\begin{equation}
\dfrac{dt}{2t}=\dfrac{dx}{x}=
\dfrac{dy}{c_1}=\dfrac{du}{-u}=\dfrac{dv}{0}.
\end{equation}
The solution of the characteristic equations yields four invariant of $X_5+c_1X_3$:
\begin{equation}
X=\dfrac{x}{\sqrt{t}},~
Y=y-\frac{c_1}{2}\log t,~\delta_1=u\sqrt{t},
~\delta_2=v.
\end{equation}
Thus, the solution of (\ref{1.12}) is given by the invariant form
\begin{equation}
u\sqrt{t}=U\left[\dfrac{x}{\sqrt{t}},y-\frac{c_1}{2}\log t\right],
v=V\left[\dfrac{x}{\sqrt{t}},y-\frac{c_1}{2}\log t\right].
\end{equation}
Then
\begin{equation}
u=\dfrac{1}{\sqrt{t}}U\left[X,Y\right],
v=V\left[X,Y\right].
\end{equation}
Based on this transformation (\ref{1.12}) reduced to
\begin{eqnarray}
(I)~~ \alpha V_{XXX}-U_{XXY}+2U U_{XY}+2U_X U_Y+&&\nonumber\\ \dfrac{1}{2}X U_{XY}+\frac{c_1}{2}U_{YY}+\frac{1}{2}U_Y&=&0,\\
V_{XX}+\beta U V_X+\frac{1}{2}X V_X+\frac{c_1}{2} V_Y&=&0.
\end{eqnarray}
For further reductions and invariant solutions, again we find the symmetries of the system $(I)$ then tabulated our results in table (\ref{T4}). The symmetries of $(I)$ are $\partial_Y$ and $\partial_V$.
\begin{table}[htb]
\caption{Solution for equation $(I)$}\label{tableExample}\label{T4}
\begin{center}
\begin{tabular}{c|c|c|c}
\hline\hline
Symmetry & Similarity variable & Reduction & Solution  \\
\hline
$\partial_Y$ & $U=G(X), $ & $H^{\prime\prime\prime}=0, $& $H=I_1+I_2X+I_3X^2$  \\
& $ V=H(X)$ & $\left(\frac{X}{2}+\beta G\right)H^\prime+H^{\prime\prime}=0$& $G=-\frac{I_2X+4I_3+2I_3X^2}{2\beta(I_2+2I_3X)}$  \\
\hline

$\partial_Y+c_3\partial_V$ & $ U=G(X),$ & $H^{\prime\prime\prime}=0, $&$H=I_1+I_2X+I_3X^2$ \\

 & $V=c_3Y+H(X)$ & $\frac{c_1c_3}{2}+\left(\frac{X}{2}+\beta G\right)H^\prime+H^{\prime\prime}=0$ &$G=-\frac{I_2X+4I_3+2I_3X^2+c_1c_3}{2\beta(I_2+2I_3X)}$ \\
\hline\hline

\end{tabular}
\end{center}
\end{table}

Next take $X_{1}+c_1X_{2}+c_2X_{3},$ then (\ref{1.12}) is reduced in to the following equation
\begin{eqnarray}
(II)~~c_2 U^{\prime\prime}-c_1c_2\left(2{U^{\prime}}^2+2UU^{\prime\prime}\right)
+{c_1}^2c_2U^{\prime\prime\prime}+\alpha{c_1}^3V^{\prime\prime\prime}&=&0,\label{1.42}\\
(1-c_1\beta U)V^{\prime}-{c_1}^2V^{\prime\prime}&=&0,\label{1.43}
\end{eqnarray}
where $u=U(r), v=V(r)$ and $r=t+c_1x+c_2y$ are invariants obtained from $X_{1}+c_1X_{2}+c_2X_{3}.$  Here $^\prime$ represents the differentiation with respect to $r$. Integrating (\ref{1.42}) twice we obtain
\begin{equation}\label{1.44}
I_2+I_1r+c_2U-c_1c_2U^2+{c_1}^2c_2U^\prime-{c_1}^3\alpha V^\prime=0.
\end{equation}
From (\ref{1.43}), we get
\begin{equation}\label{1.45}
V=I_3\int\exp\left[\int\left(\frac{1-c_1\beta U}{{c_1}^2}\right)\d r\right]\d r.
\end{equation}
Substituting (\ref{1.45}) in (\ref{1.44}) then we have
\begin{equation}\label{1.46}
U^\prime+\frac{1}{{c_1}^2}U=\frac{1}{{c_1}}U^2-\frac{I_2+I_1r}
{{c_1}^2c_2}+\frac{c_1I_3\alpha}{c_2}\exp\left[\int\left(\frac{1-c_1\beta U}{{c_1}^2}\right)\d r\right],
\end{equation}
where $I_1,I_2$ and $I_3$ are constants of integration. \\
Similarly we have performed the reduction for all other optimal system then the resultant equations are tabulated in table (\ref{T3}). The follows give the further reductions and invariant solutions for the resultant equation which are tabulated in table (\ref{T3}).

\textbf{Reductions and Solutions for $(III)$}\\
The equation $(III)$ admits the following symmetries
$$X_{11}=\partial_x, X_{12}=f_1(y)\partial_y-Vf^\prime(y)\partial_V, X_{13}=x \partial_x-U \partial_U, X_{14}= f_2(y)\partial_V $$

\begin{table}[htb]
\caption{Adjoint representation Table with $f_1=1$ and $f_2=0$ }\label{tableExample}\label{T5}
\begin{center}
\begin{tabular}{c||ccc}
\hline\hline
$\left[ Ad(e^{\epsilon X_{1i}})X_{1j}\right] $ & $X_{11}$ & $X_{12}$ & $X_{13}$  \\
\hline
$X_{11}$ & $X_{11}$ & $X_{12}$ & $X_{13}-\epsilon X_{11}$  \\
$X_{12}$ & $X_{11}$ & $X_{12}$ & $X_{13}$  \\
$X_{13}$ & $e^\epsilon X_{11}$ & $X_{12}$  & $X_{13}$ \\
\hline\hline
\end{tabular}
\end{center}
\end{table}
The table \ref{T5} shows the adjoint representation of symmetries of the resultant equation $(III)$ with conditions $f_1=1$ and $f_2=0$. The corresponding optimal system, similarity variables with respect to the optimal system, reductions and solutions are tabulated in table \ref{T6}: \begin{landscape}
\begin{table}[htb]
\caption{Solution for equation $(III)$}\label{tableExample}\label{T6}
\begin{center}
\begin{tabular}{c|c|c|c}
\hline\hline
Optimal & Similarity  & Reduction & Solution  \\
system&  variable & & \\
\hline
$X_{11}$& $U(x,y)=G(y), $ & & $U(x,y)=G(y)$  \\
& $ V(x,y)=H(y)$ & & $ V(x,y)=H(y)$   \\
\hline

$X_{12}$  & $ U(x,y)=G(x),$ & $H^{\prime\prime\prime}=0, $&$H=I_1+I_2x+I_3x^2$ \\

 & $V(x,y)=H(x)$ & $\beta G H^\prime+H^{\prime\prime}=0$ &$G=-\frac{2I_3}{\beta(I_2+2I_3x)}$ \\
 \hline
$X_{13}$& $U(x,y)=\frac{G(y)}{x}, $ & $(1+2G)G^\prime=0, $& $G(y)=I_1$  \\
& $ V(x,y)=H(y)$ & & H(y)=H(y)  \\
\hline

$X_{11}+c_1 X_{12}$  & $r=y-c_1x$,  & $2({G^\prime}^2+GG^{\prime\prime})+ $&$H=I_1\int\exp\left[\frac{\beta}{c_1}\int G dr\right]dr+I_2$ \\

 & $U(x,y)=G(r),$ & $c_1G^{\prime\prime\prime}+{c_1}^2\alpha H^{\prime\prime\prime}=0,$ &$c_1G^\prime =I_4+I_3r-G^2-$ \\
 &$V(x,y)=H(r)$ & $\beta G H^\prime-c_1H^{\prime\prime}=0$ &${c_1}^2\alpha I_1 \exp\left[\frac{\beta}{c_1}\int G dr\right]$ \\
 \hline
 $X_{13}+c_1 X_{12}$ &  $r=y-c_1x$,  & ${c_1}^2G^{\prime\prime\prime}+3c_1G^{\prime\prime}+2G^\prime
+ $& $H=I_1\int\exp\left[-\frac{1}{c_1}\int(1-\beta G )dr\right]dr+I_2$  \\
& $U(x,y)=\frac{G(r)}{x}$  & $4GG^\prime+2c_1({G^\prime}^2+GG^{\prime\prime})+$& $(**)~~{c_1}^2G^{\prime\prime}+3c_1G^{\prime}+2G =$  \\
& $V(x,y)=H(r)$ & $2c_1\alpha H+3{c_1}^2\alpha H^{\prime\prime}+{c_1}^3\alpha H^{\prime\prime\prime}=0$& $I_3-2G^2-2GG^\prime-$  \\
&  & $(1-\beta G )H^\prime-c_1H^{\prime\prime}=0$& $(G+\beta G^2+c_1G^\prime)I_1c_1\alpha\beta\exp\left[-\frac{1}{c_1}\int(1-\beta G) dr\right]$  \\
\hline\hline

\end{tabular}
\end{center}
 The trivial symmetry of equation $(**)$ is $\partial_r$ and the corresponding canonical variable are $G^\prime=W(G).$ Based on these transformation the equation $(**)$ reduced in to the following
\begin{eqnarray}
{c_1}^2WW^\prime+3c_1W+2G &=&I_3-2G^2-2GW-\nonumber\\&&I_1c_1\alpha\beta(G+\beta G^2+c_1W)\exp\left[-\frac{1}{c_1}\int(1-\beta G) dr\right],\label{3}
\end{eqnarray}
where $^\prime$ represents the differentiation with respect to $G$.
\end{table}
\end{landscape}
\textbf{Reductions and Solutions for $(IV)$}\\
The solution of system $(IV)$ is given by
\begin{eqnarray}
U(t,y)&=&\int f(t)dt+g(y),\\
V(t,y)&=&h(y).
\end{eqnarray}

\textbf{Reductions and Solutions for $(V)$}\\
The solution of system $(V)$ is given by
\begin{eqnarray}
U(t,x)&=&\frac{f^\prime(t)+g^\prime(t)x+h^\prime(t)x^2-2h(t)}
{\beta(g(t)+2xh(t))},\\
V(t,y)&=&f(t)+g(t)x+h(t)x^2.
\end{eqnarray}

\textbf{Reductions and Solutions for $(VI)$}\\

The trivial symmetry of the system $(VI)$ is $\partial_Y.$ Based on the symmetry the similarity variables are $U(X,Y)=G(X)$ and $V(X,Y)=H(X)$ and then the reduced systen is given by
 \begin{eqnarray}
H_{XXX}&=&0,\\
(X+2\beta G)H_X+2H_{XX}&=&0.
\end{eqnarray}
The solution of above system are given by
\begin{eqnarray}
H&=&I_1+I_2X+I_3X^2,\\
G&=&-\frac{4I_3+I_2X+2I_3X^2}{2\beta(I_2+2I_3X)},
\end{eqnarray}
where $I_1,I_2$ and $I_3$ are constants of integration.


\subsection{Case $m=1$ and $n=1$}
If $m=1$ and $n=1$ then the system (\ref{1}) becomes
\begin{eqnarray}\label{1.13}
u_{ty}-\left( u-u_{x}\right) _{xy}-\alpha v_{xxx} &=&0, \nonumber\\
v_{t}-v_{xx}-\beta u v_{x} &=&0.
\end{eqnarray}
The above system admitted the following Lie symmetries
\begin{eqnarray*}
Y_{1} &=&\partial _{t}~,~Y_{2}=\partial _{x}~,~Y_{3}=\psi(y)\partial
_{y}-\psi^\prime v \partial_v~,~Y_{4}^{\infty }=\phi \left( y\right) \partial _{v}~, \\
Y_{5} &=&2t \partial _{t}+\left( x-t\right) \partial _{x}+\left( \frac{1-\beta u}{\beta}\right) \partial _{u}.
\end{eqnarray*}
\begin{table}[htb]
\caption{Commutator Table for case 1.2}\label{tableExample}\label{T7}
\begin{center}
\begin{tabular}{c||ccccc}
\hline\hline
$\left[ Y_{I},Y_{J}\right] $ & $Y_{1}$ & $Y_{2}$ & $Y_{3}$ & $Y_{4}$ & $Y_{5}$ \\
\hline
$Y_{1}$ & $0$ & $0$ & $0$ & $0$& $(2Y_{1}-Y_2)$  \\
$Y_{2}$ & $0$ & $0$ & $0$  & $0$& $Y_{2}$\\
$Y_{3}$ & $0$ & $0$  & $0$& $(\psi^{\prime}\phi+\phi^{\prime}\psi)\partial_v$ & $0$\\
$Y_{4}$ & $0$ & $0$  & $-(\psi^{\prime}\phi+\phi^{\prime}\psi)\partial_v$& $0$ & $0$\\
$Y_{5}$ & $-2Y_{1}+Y_2$ & $-Y_{2}$  & $0$& $0$ & $0$\\
\hline\hline
\end{tabular}
\end{center}
\end{table}

\begin{table}[htb]
\caption{Adjoint representation Table for case 1.2 }\label{tableExample}\label{T8}
\begin{center}
\begin{tabular}{c||cccc}
\hline\hline
$\left[ Ad(e^{\epsilon Y_i})Y_j\right] $ & $Y_{1}$ & $Y_{2}$ & $Y_{3}$ & $Y_{5}$ \\
\hline
$Y_{1}$ & $Y_{1}$ & $Y_{2}$ & $Y_{3}$ & $Y_{5}-2\epsilon Y_{1}+\epsilon Y_2$  \\
$Y_{2}$ & $Y_{1}$ & $Y_{2}$ & $Y_{3}$  & $Y_{5}-\epsilon Y_{2}$\\
$Y_{3}$ & $Y_{1}$ & $Y_{2}$  & $Y_{3}$ & $Y_{5}$\\
$Y_{5}$ & $e^{2\epsilon}Y_{1}
-e^{\epsilon}(e^{\epsilon}-1)Y_{2}$ & $e^{\epsilon}Y_{2}$  & $Y_{3}$& $Y_{5}$\\
\hline\hline
\end{tabular}
\end{center}
\end{table}
Table \ref{T7}
and table \ref{T8} give the commutator relation and adjoint representation, respectively, for the symmetries of system (\ref{1.13}). Indeed, the adjoint representation tabulated with condition $\psi(y)=1$ and $\phi(y)=0.$
\begin{landscape}
\begin{table}[htb]
\caption{Reductions for the case $m=1,n=1$ }\label{tableExample}\label{T9}
\begin{center}
\begin{tabular}{c|c|cc}
\hline\hline
Optimal System & Similarity variable & &Reductions  \\
\hline
$ Y_1+c_1Y_2$ & $X=x-c_1 t,~Y=y,$ & $(VII)$&$U_{YXX}-(1+c_1)U_{YX}-\alpha v_{XXX} =0$  \\

 & $u=U[X,Y],~v=V[X,Y]$ &  &$V_{XX}+(c_1+\beta U)V_{X} =0$\\
\hline

$Y_5+c_1Y_3$ & $X=\frac{1}{\sqrt{t}}(x+t),~Y=y-\frac{c_1}{2}\log t,$ & $(VIII)$&$U_Y+c_1U_{YY}+X U_{YX}-2U_{YXX}+2\alpha v_{XXX} =0 $ \\

 &u=U[X,Y],~v=V[X,Y]  & &$c_1 V_Y+(X+2\beta U)V_{X}+2V_{XX} =0$ \\
\hline
$Y_1+c_1 Y_2 +c_2 Y_3$ & $r=t+c_1x+c_2y$ & $(IX)$&$c_2 U^{\prime\prime}-c_1c_2U^{\prime\prime}
+{c_1}^2c_2U^{\prime\prime\prime}-\alpha{c_1}^3V^{\prime\prime\prime}=0$
 \\

&u=U(r), v=V(r) & &$(1-c_1\beta U)V^{\prime}-{c_1}^2V^{\prime\prime}=0$   \\
\hline
$Y_1$ & $u(t,x,y)=U(x,y),v(t,x,y)=V(x,y)$ & $(X)$&$U_{xy}-U_{xxy}+\alpha V_{xxx}=0, $ \\
& & &$\beta U V_x+V_{xx}=0 $  \\ \hline
$Y_2$ & $u(t,x,y)=U(t,y), v(t,x,y)=V(t,y)$ & $(XI)$& $U_{ty}=0,
$  \\
 & & &$V_t=0$  \\ \hline
$Y_3$ & $u(t,x,y)=U(t,x), v(t,x,y)=V(t,x),$ & $(XII)$&$\beta U V_x+V_{xx}-V_t=0$  \\
 &  & &$V_{xxx}=0$  \\ \hline
 $Y_5$& $X=\frac{t+x}{\sqrt t}, Y=y,$ & $(XIII)$&$2\alpha V_{XXX}+ U_Y+X U_{XY}-2U_{YXX}=0$  \\
  &$u(t,x,y)=\frac{1}{\beta}+\frac{U(X,Y)}{\sqrt t}, v(t,x,y)=V(X,Y)$ &  &$(X+2\beta U)V_X+2V_{XX}=0$ \\ \hline\hline
\end{tabular}
\end{center}
\end{table}
\end{landscape}
Now we have examined the optimal system, similarity variables for the optimal system, the resultant equation based on the similarity variables are tabulated in table \ref{T9}. The follows, giving the reductions and solutions for the resultant equations which are tabulated in table \ref{T9}.

\textbf{Reductions and solutions for $(VII)$ }\\
%
 The resultant system $(VII)$ have the symmetries $\partial_X,~f_1(Y)\partial_Y-V{f_1}^{\prime}(Y)\partial_V,~f_2(Y)\partial_V$. If $f_1(Y)=f_2(Y)=1$ then the optimal system, corresponding similarity variables, reductions and solutions of $(VII)$ tabulated in table \ref{T10}.
 \begin{table}[htb]
\caption{Solution for equation $(VII)$}\label{tableExample}\label{T10}
\begin{center}
\begin{tabular}{c|c|c|c}
\hline\hline
Optimal & Similarity  & Reduction & Solution  \\
system&  variable & & \\
\hline
$\partial_X+c_2\partial_Y$& $r=X-c_2Y, $ & $c_2(1+c_1)G^{\prime\prime}-c_2G^{\prime\prime\prime}-\alpha H^{\prime\prime\prime}=0$& $H=I_1\int\exp\left[-\int(c_1+\beta G)dr\right]dr+I_2$  \\
& $U(X,Y)=G(r), $ &$(c_1+\beta G)H^\prime+H^{\prime\prime}=0$ & $G^\prime-(1+c_1) G=I_3 r+I_4-$  \\
& $ V(X,Y)=H(r)$ & & $\alpha c_2 I_1 \exp\left[-\int(c_1+\beta G)dr\right]$   \\
\hline

$\partial_Y+c_3\partial_V$  & $U(X,Y)=G(X),$ & $H^{\prime\prime\prime}=0, $&$H=I_1+I_2X+I_3X^2$ \\

 & $V(X,Y)=\frac{Y}{c_3}+H(X)$ & $(c_1+\beta G) H^\prime+H^{\prime\prime}=0$ &$G=-\frac{c_1I_2+2I_3+2c_1I_3X}{\beta(I_2+2I_3X)}$ \\
 \hline
$\partial_Y$& $U(X,Y)=G(X), $ & $H^{\prime\prime\prime}=0, $& $H=I_1+I_2X+I_3X^2$  \\
& $ V(X,Y)=H(X)$ & $(c_1+\beta G)H^\prime+H^{\prime\prime}=0$& $G=-\frac{c_1I_2+2I_3+2c_1I_3X}{\beta(I_2+2I_3X)}$  \\
\hline

 \hline\hline

\end{tabular}
\end{center}
\end{table}

%
\textbf{Reductions and solutions for $(VIII)$ }\\
The resultant system $(VIII)$ have the symmetries $\partial_Y,~\partial_V$. The optimal system, corresponding similarity variables, reductions and solutions of $(VIII)$ tabulated in table \ref{T11}.

\begin{table}[htb]
\caption{Solution for equation $(VIII)$}\label{tableExample}\label{T11}
\begin{center}
\begin{tabular}{c|c|c|c}
\hline\hline
Symmetry & Similarity variable & Reduction & Solution  \\
\hline
$\partial_Y$ & $U=G(X), $ & $H^{\prime\prime\prime}=0, $& $H=I_1+I_2X+I_3X^2$  \\
& $ V=H(X)$ & $\left(\frac{X}{2}+\beta G\right)H^\prime+H^{\prime\prime}=0$& $G=-\frac{I_2X+4I_3+2I_3X^2}{2\beta(I_2+2I_3X)}$  \\
\hline

$\partial_Y+c_3\partial_V$ & $ U=G(X),$ & $H^{\prime\prime\prime}=0, $&$H=I_1+I_2X+I_3X^2$ \\

 & $V=\frac{Y}{c_3}+H(X)$ & $\frac{c_1}{2c_3}+\left(\frac{X}{2}+\beta G\right)H^\prime+H^{\prime\prime}=0$ &$G=-\frac{c_1+c_3(I_2X+4I_3+2I_3X^2)}{2c_3\beta(I_2+2I_3X)}$ \\
\hline\hline

\end{tabular}
\end{center}
\end{table}


%
%

\textbf{Reductions and solutions for $(IX)$ }\\
Integrating the resultant equation $(IX)$ we obtain the following reduced equation
\begin{eqnarray}\label{R11}
V&=&I_3\int\exp\left[\int\left(\frac{1-c_1\beta U}{{c_1}^2}\right)\d r\right]\d r,\\
U^\prime+\frac{1}{{c_1}^2}U&=&\frac{1}{{c_1}}U-\frac{I_2+I_1r}
{{c_1}^2c_2}+\frac{c_1I_3\alpha}{c_2}\exp\left[\int\left(\frac{1-c_1\beta U}{{c_1}^2}\right)\d r\right],
\end{eqnarray}
where $I_1,I_2$ and $I_3$ are constants of integration.\\

\textbf{Reductions and solutions for $(X)$ }\\
The system $(X)$ admits the symmetries
$Y_{11}=\partial_x, Y_{12}=f_1(y)\partial_y-Vf^\prime(y)\partial_V, Y_{13}= f_2(y)\partial_V .$ The adjoint representation given by the table \ref{T12} and corresponding optimal system, similarity variables, reductions and solutions are given in table \ref{T13}.
\begin{table}[htb]
\caption{Adjoint representation Table with $f_1=1$ and $f_2=0$ }\label{tableExample}\label{T12}
\begin{center}
\begin{tabular}{c||cc}
\hline\hline
$\left[ Ad(e^{\epsilon Y_{1i}})Y_{1j}\right] $ & $Y_{11}$ & $Y_{12}$\\
\hline
$Y_{11}$ & $Y_{11}$ & $Y_{12}$  \\
$Y_{12}$ & $Y_{11}$ & $Y_{12}$   \\

\hline\hline
\end{tabular}
\end{center}
\end{table}

\begin{table}[htb]
\caption{Solution for equation $(X)$}\label{tableExample}\label{T13}
\begin{center}
\begin{tabular}{c|c|c|c}
\hline\hline
Optimal & Similarity  & Reduction & Solution  \\
system&  variable & & \\
\hline
$Y_{11}$ & $U(x,y)=G(x), $ & & $U(x,y)=G(x)$  \\
& $ V(x,y)=H(x)$ & & $V(x,y)=H(x)$   \\
\hline

$Y_{12}$  & $U(x,y)=G(x),$ & $H^{\prime\prime\prime}=0, $&$H=I_1+I_2x+I_3x^2$ \\

 & $V(x,y)=H(x)$ & $\beta G H^\prime+H^{\prime\prime}=0$ &$G=-\frac{2I_3}{\beta(I_2+2I_3x)}$ \\
 \hline
$Y_{11}+c_1 Y_{12}$& $r=y-c_1x$, & ${c_1}G^{\prime\prime\prime}+G^{\prime\prime}+{c_1}^2\alpha H^{\prime\prime\prime}=0, $& $H=I_1\int\exp\left[\frac{\beta}{c_1}\int G dr\right]dr+I_2$  \\
& $U(x,y)=\frac{G(r)}{x}$, & $(\beta G )H^\prime-c_1H^{\prime\prime}=0 $& $c_1G^{\prime}+G=I_3r+I_4-$  \\
& $ V(x,y)=H(r)$ & & ${c_1}^2I_1\alpha\exp\left[\frac{\beta}{c_1}\int G dr\right]$  \\
 \hline\hline

\end{tabular}
\end{center}
\end{table}

%

%

\textbf{Reductions and solutions for $(XI)$ }\\
The solution of the system $(XI)$ is given by
\begin{eqnarray}
U(t,y)&=&\int f(t)dt+g(y),\\
V(t,y)&=&h(y).
\end{eqnarray}
\textbf{Reductions and solutions for $(XII)$ }\\
%
The solution of the system $(XII)$ is given by
\begin{eqnarray}
U(t,x)&=&\frac{f^\prime(t)+g^\prime(t)x+h^\prime(t)x^2-2h(t)}
{\beta(g^\prime(t)+2xh(t))},\\
V(t,y)&=&f(t)+g(t)x+h(t)x^2.
\end{eqnarray}

\textbf{Reductions and solutions for $(XIII)$ }\\
%
The trivial symmetry of the system $(XIII)$ is $\partial_Y.$ Based on the symmetry the similarity variable is $U(X,Y)=G(X)$ and $V(X,Y)=H(X)$ and the reduced equation is given by
 \begin{eqnarray}
H_{XXX}&=&0,\\
(X+2\beta G)H_X+2H_{XX}&=&0.
\end{eqnarray}
The solution of above system are given by
\begin{eqnarray}
H&=&I_1+I_2X+I_3X^2,\\
G&=&-\frac{4I_3+I_2X+2I_3X^2}{2\beta(I_2+2I_3X)},
\end{eqnarray}
where $I_1,I_2$ and $I_3$ are constants of integration.


\subsection{Case $n\neq 1$,~$m=n-1$}
If $m=n-1$ and $n\ne1$ then the system (\ref{1}) yields the following equation
\begin{eqnarray}\label{1.14}
u_{ty}-\left( u^{n}-u_{x}\right) _{xy}-\alpha v_{xxx} &=&0 ,\nonumber\\
v_{t}-v_{xx}-\beta u^{n-1} v_{x} &=&0.
\end{eqnarray}

The above system admitted Lie symmetries as we are given below:
\begin{eqnarray*}
Z_{1} &=&\partial _{t}~,~Z_{2}=\partial _{x}~,~Z_{3}=\psi(y)\partial
_{y}-\psi^\prime v \partial_v~,~Z_{4}^{\infty }=\phi \left( y\right) \partial _{v}~, \\
Z_{5} &=&2t \partial _{t}+x \partial
_{x}+\left(\frac{1}{1-n}\right)u\partial _{u}+\left(\frac{2-n}{1-n}\right)v\partial _{v}.
\end{eqnarray*}

The commutator table for the symmetries are displayed in table \ref{T14}. Table \ref{T15} gives adjoint representation of the symmetries with condition $\psi(y)=1$ and $\phi(y)=0.$ Optimal system, similarity variables and corresponding resultant equations are given in table \ref{T16}.

\begin{table}[htb]
\caption{Commutator Table for case 1.3}\label{tableExample}\label{T14}
\begin{center}
\begin{tabular}{c||ccccc}
\hline\hline
$\left[ Z_{I},Z_{J}\right] $ & $Z_{1}$ & $Z_{2}$ & $Z_{3}$ & $Z_{4}$ & $Z_{5}$ \\
\hline
$Z_{1}$ & $0$ & $0$ & $0$ & $0$& $2Z_{1}$  \\
$Z_{2}$ & $0$ & $0$ & $0$  & $0$& $ Z_{2}$\\
$Z_{3}$ & $0$ & $0$  & $0$& $(\psi^{\prime}\phi+\phi^{\prime}\psi)\partial_v$ & $0$\\
$Z_{4}$ & $0$ & $0$  & $-(\psi^{\prime}\phi+\phi^{\prime}\psi)\partial_v$& $0$ & $\left(\frac{n-2}{n-1}\right)Z_{4}$\\
$Z_{5}$ & $-2Z_{1}$ & $- Z_{2}$  & $0$& $-\left(\frac{n-2}{n-1}\right)Z_{4}$ & $0$\\
\hline\hline
\end{tabular}
\end{center}
\end{table}

\begin{table}[htb]
\caption{Adjoint representation Table for case 1.3 }\label{tableExample}\label{T15}
\begin{center}
\begin{tabular}{c||cccc}
\hline\hline
$\left[ Ad(e^{\epsilon Z_i})Z_j\right] $ & $Z_{1}$ & $Z_{2}$ & $Z_{3}$ & $Z_{5}$ \\
\hline
$Z_{1}$ & $Z_{1}$ & $Z_{2}$ & $Z_{3}$ & $Z_{5}-2\epsilon Z_{1}$  \\
$Z_{2}$ & $Z_{1}$ & $Z_{2}$ & $Z_{3}$  & $Z_{5}-\epsilon Z_{2}$\\
$Z_{3}$ & $Z_{1}$ & $Z_{2}$  & $Z_{3}$ & $Z_{5}$\\
$Z_{5}$ & $e^{2\epsilon}Z_{1}$ & $e^{\epsilon}Z_{2}$  & $Z_{3}$& $Z_{5}$\\
\hline\hline
\end{tabular}
\end{center}
\end{table}

\begin{landscape}
\begin{table}[htb]
\caption{Reductions for the case $m=n-1,n\ne1$ }\label{tableExample}\label{T16}
\begin{center}
\begin{tabular}{c|c|cc}
\hline\hline
Optimal System & Similarity variable & &Reductions  \\
\hline
$Z_5+c_1Z_3$ & $X=\dfrac{x}{\sqrt{t}},~
Y=y-\frac{c_1}{2}\log t,$ & $(XIV)$&$\alpha V_{XXX}-U_{XXY}+nU^{n-1} U_{XY}+\dfrac{1}{2}X U_{XY}+$  \\

 & $u=t^{\frac{1}{2(1-n)}}U\left[X,Y\right],
v=t^{\frac{2-n}{2(1-n)}}V\left[X,Y\right]$ &  &$n(n-1)U^{n-2}U_X U_Y+\frac{c_1}{2}U_{YY}+\frac{1}{2(n-1)}U_Y=0$\\

 &  &  &$V_{XX}+\beta U^{n-1} V_X+\dfrac{1}{2}X V_X+\frac{c_1}{2}V_Y-\left(\frac{n-2}{2(n-1)}\right)V=0$ \\
\hline

$Z_1+c_1Z_2+c_2Z_3$ & $r=t+c_1x+c_2y, $ & $(XV)$&$c_2 U^{\prime\prime}-c_1c_2\left(n(n-1)U^{n-2}{U^{\prime}}^2+nU^{n-1}U^{\prime\prime}\right)
+{c_1}^2c_2U^{\prime\prime\prime}+\alpha{c_1}^3V^{\prime\prime\prime}=0
$ \\

 & $u=U(r), v=V(r)$ & &$(1-c_1\beta U^{n-1})V^{\prime}-{c_1}^2V^{\prime\prime}=0$ \\
\hline
$Z_{1}$ & $u=U[x,y],v=V[x,y]$ & $(XVI)$&$n(n-1)U^{n-2}U_x U_y+nU^{n-1}U_{xy}-U_{xxy}+\alpha V_{xxx}=0$
 \\

& & &$\beta U^{n-1} V_x+V_{xx}=0$   \\
\hline
$Z_{2}$ & $u(t,x,y)=U(t,y), v(t,x,y)=V(t,y)$ & $(XVII)$&$U_{ty}=0, $ \\
& & &$V_t=0 $  \\ \hline
$Z_{3}$ & $u(t,x,y)=U(t,x), v(t,x,y)=V(t,x)$ & $(XVIII)$&$\beta U^{n-1} V_x+V_{xx}-V_t=0
$  \\
 & & &$V_{xxx}=0$  \\ \hline
$Z_{5}$ & $x=X \sqrt t, y=Y,$ & $(XIX)$&${A}^{\frac{3}{2}}{B}^{\frac{1}{2}}U_y+n{A}^{\frac{n}{2}}
{B}^{\frac{n}{2}}((n^2+1)A-n{A}^n)U^{n-2}U_X U_Y+$  \\
 & $u(t,x,y)=(2(1-n)t)^{\frac{1}{2(1-n)}}U(X,Y), $  & &$(n-1){A}^{\frac{3}{2}}{B}^{\frac{1}{2}}X U_{XY}+n(n-1){A}^{\frac{n+2}{2}}{B}^{\frac{n}{2}}
U^{n-1}U_{XY}+$  \\
&$v(t,x,y)=(2(n-1)t)^{\frac{n-2}{2(n-1)}}V(X,Y)$ & & ${A}^{\frac{3n}{2}}{B}^{\frac{4-3n}{2}}\alpha V_{XXX}+{A}^{\frac{1+2n}{2}}{B}^{\frac{3-2n}{2}}
U_{XXY}=0 $ \\
& & &$(n-2)V-(n-1)XV_X+{(2(1-n))}^{\frac{1}{2}}\beta U^{n-1}V_X-2(n-1)V_{XX}=0$  \\ \hline\hline
\end{tabular}
\end{center}
where $A=2^{\frac{1}{n-1}}$ and $B={(1-n)^{\frac{1}{(1-n)}}}.$
\end{table}
\end{landscape}

\begin{landscape}
\begin{table}[htb]
\caption{Solution for equation $(XIV)$}\label{T17}
\begin{center}
\begin{tabular}{c|c|c|c}
\hline\hline
Symmetry & Similarity variable & Reduction & Solution  \\
\hline
$\partial_Y$ & $U=G(X), $ & $H^{\prime\prime\prime}=0, $&$H=I_1+I_2X+I_3X^2$ \\
& $ V=H(X)$ & $\frac{(n-2)}{(n-1)}H-(X+2\beta G^{n-1})H^\prime-2H^{\prime\prime}=0$& $G=\left[-\frac{2(n-1)\beta (I_2+2I_3X)}{(2-n)I_1+I_2X+4(n-1)I_3+nI_3X^2}\right]^{\frac{1}{n-1}}$ \\
\hline

$\partial_Y+c_3e^{\frac{(n-2)Y}{(n-1)c_1}}\partial_V$ & $ U=G(X),$ & $H^{\prime\prime\prime}=0, $&$H=I_1+I_2X+I_3X^2$ \\

 & $V=\frac{(n-1)c_1}{c_3(n-2)}e^{\frac{(n-2)Y}{(n-1)c_1}}+H(X)$ & $\frac{(n-2)}{(n-1)}H-(X+2\beta G^{n-1})H^\prime-2H^{\prime\prime}=0$ &$G=\left[-\frac{2(n-1)\beta (I_2+2I_3X)}{(2-n)I_1+I_2X+4(n-1)I_3+nI_3X^2}\right]^{\frac{1}{n-1}}$ \\
\hline\hline

\end{tabular}
\end{center}
\end{table}
\end{landscape}

%

\textbf{Reductions and solutions for $(XIV)$ }\\

%
The system $(XIV)$ has two symmetries which are given as $\partial_Y$ and $e^{\frac{(n-2)Y}{(n-1)c_1}}\partial_V.$ Corresponding reductions and solutions are tabulated in table \ref{T17}.

\textbf{Reductions and solutions for $(XV)$ }\\
The resultant equation further reduced to the following equations
\begin{eqnarray}\label{1.55}
V&=&I_3\int\exp\left[\int\left(\frac{1-c_1\beta U^{n-1}}{{c_1}^2}\right)\d r\right]\d r,\\
U^\prime+\frac{1}{{c_1}^2}U&=&\frac{1}{{c_1}}U^n-\frac{I_2+I_1r}
{{c_1}^2c_2}+\frac{c_1I_3\alpha}{c_2}\exp\left[\int\left(\frac{1-c_1\beta U^{n-1}}{{c_1}^2}\right)\d r\right],
\end{eqnarray}
where $I_1,I_2$ and $I_3$ are constants of integration.\\

\textbf{Reductions and solutions for $(XVI)$ }\\
The equation $(XVI)$ admits the following symmetries
$$Z_{11}=\partial_x, Z_{12}=f_1(y)\partial_y-Vf^\prime(y)\partial_V, Z_{13}=x \partial_x-\left(\frac{1}{n-1}\right)U \partial_U+\left(\frac{n-2}{n-1}\right)V\partial_V, Z_{14}= f_2(y)\partial_V .$$ The adjoint representation with conditions $f_1(y)=1$ and $f_2(y)=0$ is tabulated in table \ref{T18}
and corresponding optimal system, similarity variables, reduced system and solutions are given by table \ref{T19}.
\begin{table}[htb]
\caption{Adjoint representation Table with $f_1=1$ and $f_2=0$ }\label{tableExample}\label{T18}
\begin{center}
\begin{tabular}{c||ccc}
\hline\hline
$\left[ Ad(e^{\epsilon Z_{1i}})Z_{1j}\right] $ & $Z_{11}$ & $Z_{12}$ & $Z_{13}$  \\
\hline
$Z_{11}$ & $Z_{11}$ & $Z_{12}$ & $Z_{13}-\epsilon Z_{11}$  \\
$Z_{12}$ & $Z_{11}$ & $Z_{12}$ & $Z_{13}$  \\
$Z_{13}$ & $e^\epsilon Z_{11}$ & $Z_{12}$  & $Z_{13}$ \\
\hline\hline
\end{tabular}
\end{center}
\end{table}

\begin{landscape}
\begin{table}[htb]
\caption{Solution for equation $(XVI)$}\label{tableExample}\label{T19}
\begin{center}
\begin{tabular}{c|c|c|c}
\hline\hline
Optimal & Similarity  & Reduction & Solution  \\
system&  variable & & \\
\hline
$Z_{11}$& $U(x,y)=G(y), $ & & $U(x,y)=G(y)$  \\
& $ V(x,y)=H(y)$ & & $ V(x,y)=H(y)$   \\
\hline

$Z_{12}$  & $ U(x,y)=G(x),$ & $H^{\prime\prime\prime}=0, $&$H=I_1+I_2x+I_3x^2$ \\

 & $V(x,y)=x^{\frac{n-2}{n-1}}H(y)$ & $\beta G^{n-1} H^\prime+H^{\prime\prime}=0$ &$G=\left[-\frac{2I_3}{\beta(I_2+2I_3x)}\right]^{\frac{1}{n-1}}$ \\
 \hline
$Z_{13}$& $U(x,y)=(x-nx)^{\frac{1}{1-n}}G(y), $ & $(2-n)\alpha G H-(1-n)^{\frac{2-n}{1-n}}GG^\prime + $& $G(y)={(-\beta)}^{\frac{1}{1-n}}$  \\
&  & $n(1-n)^{\frac{n-2}{n-1}}G^nG^\prime=0, $& \\
& $ V(x,y)=H(y)$ & $
G+\beta G^n=0$ & $H(y)=0$  \\
\hline

$Z_{11}+c_1 Z_{12}$  & $r=y-c_1x$,  & $n(n-1)(G^{n-2}{G^\prime}^2)+nG^{n-1}G^{\prime\prime}
+ $&$H=I_1\int\exp\left[\frac{\beta}{c_1}\int G^{n-1} dr\right]dr+I_2$ \\

 & $U(x,y)=G(r),$ & $c_1G^{\prime\prime\prime}+{c_1}^2\alpha H^{\prime\prime\prime}=0,$ &$c_1G^\prime =I_4+I_3r-G^n-$ \\
 &$V(x,y)=H(r)$ & $\beta G^{n-1} H^\prime-c_1H^{\prime\prime}=0$ &${c_1}^2\alpha I_1 \exp\left[\frac{\beta}{c_1}\int G^{n-1} dr\right]$ \\
 \hline
 $Z_{13}+c_1 Z_{12}$ &  $r=y-c_1x$,  & $n\left(\frac{n-2}{n-1}\right)\alpha H+n^2(1-n)^{\frac{1}{1-n}}G^{n-1}G^{\prime}
+$&   \\
  & $U(x,y)=x^{\frac{1}{1-n}}(1-n)^{\frac{1}{1-n}}G(r)$   & $c_1n(1-n)^{\frac{3-2n}{1-n}}G^{n-2}{G^\prime}^2+c_1(n^2-2n-2)\alpha H^\prime+ $&   \\
& $V(x,y)=x^{\frac{n-2}{n-1}}H(r)$  & $c_1(1+n)(1-n)^{\frac{2-n}{1-n}}G^{\prime\prime}
-c_1n(1-n)^{\frac{2-n}{1-n}}G^{n-1}G^{\prime\prime}-$&   \\

  &  & $3{c_1}^2(n-1)\alpha H^{\prime\prime}-{c_1}^2(1-n)^{\frac{3-2n}{1-n}}G^{\prime\prime\prime}-$& \\

&  & $n(1-n)^{\frac{1}{1-n}}{G^\prime}-{ c_1}^3(n-1)^2\alpha H^{\prime\prime\prime}=0$&  \\
&  & $\left(\frac{n-2}{n-1}\right)(1+\beta G^{n-1} )H+c_1(n-3)H^{\prime}-$&   \\

&  & $c_1\beta G^{n-1}H^\prime-{c_1}^2(n-1)H^{\prime\prime}=0$& \\
\hline\hline

\end{tabular}
\end{center}
\end{table}
\end{landscape}

\textbf{Reductions and solutions for $(XVII)$ }\\
The solution of the system $(XVII)$ is given by
\begin{eqnarray}
U(t,y)&=&\int f(t)dt+g(y),\\
V(t,y)&=&h(y).
\end{eqnarray}

\textbf{Reductions and solutions for $(XVIII)$ }\\

The solution of the system $(XVIII)$ is given by
\begin{eqnarray}
U(t,x)&=&\left[\frac{f^\prime(t)+g^\prime(t)x
+h^\prime(t)x^2-2h(t)}{\beta(g(t)+2xh(t))}\right]^{\frac{1}{n-1}},\\
V(t,y)&=&f(t)+g(t)x+h(t)x^2.
\end{eqnarray}

\textbf{Reductions and solutions for $(XIX)$ }\\

The trivial symmetry of the system $(XIX)$ is $\partial_Y.$ Based on this symmetry the system is reduced in to
\begin{eqnarray}
 H^{\prime\prime\prime}=0,&&\\
(n-2)H-(n-1)XH^\prime+&&\nonumber\\{(2(1-n))}^{\frac{1}{2}}\beta G^{n-1}H^\prime-2(n-1)H^{\prime\prime}=0.&&
\end{eqnarray}
The solution of above system are given by
\begin{eqnarray}
H&=&I_1+I_2X+I_3X^2,\\
G&=&\left[\frac{{(2(1-n))}^{\frac{1}{2}}\beta (I_2+2I_3 X)}{(2-n)I_1+I_2X+4(n-1)I_3+n I_3X^2}\right]^{\frac{1}{1-n}}.
\end{eqnarray}


\subsection{Arbitrary $n,$ $m$}
For arbitrary $m$ and $n$:

The system (\ref{1}) admitted Lie symmetries as follows
\[
\Gamma _{1}=\partial _{t}~,~\Gamma _{2}=\partial _{x}~,~\Gamma _{3}=\partial
_{y}~,~\Gamma _{4}^{\infty }=\psi \left( y\right) \partial _{v}
\]

The commutator table for the symmetries are displayed in table \ref{T20}. Table \ref{T21} gives adjoint representation of the symmetries with condition $\psi(y)=0$. Optimal system, similarity variables and corresponding resultant equations are given in table \ref{T22}.
\begin{table}[htb]
\caption{Commutator Table for case 1.4}\label{tableExample}\label{T20}
\begin{center}
\begin{tabular}{c||cccc}
\hline\hline
$\left[ \Gamma_{I},\Gamma_{J}\right] $ & $\Gamma_{1}$ & $\Gamma_{2}$ & $\Gamma_{3}$ & $\Gamma_{4}$ \\
\hline
$\Gamma_{1}$ & $0$ & $0$ & $0$ & $0$  \\
$\Gamma_{2}$ & $0$ & $0$ & $0$  & $0$\\
$\Gamma_{3}$ & $0$ & $0$  & $0$& $\Gamma_{4}$ \\
$\Gamma_{4}$ & $0$ & $0$  & $-\Gamma_{4}$& $0$ \\
\hline\hline
\end{tabular}
\end{center}
\end{table}

\begin{table}[htb]
\caption{Adjoint representation table for case 1.4}\label{tableExample}\label{T21}
\begin{center}
\begin{tabular}{c||ccc}
\hline\hline
$\left[ Ad(e^{\epsilon \Gamma_{I}})\Gamma_{J}\right] $ & $\Gamma_{1}$ & $\Gamma_{2}$ & $\Gamma_{3}$ \\
\hline
$\Gamma_{1}$ & $\Gamma_{1}$ & $\Gamma_{2}$ & $\Gamma_{3}$   \\
$\Gamma_{2}$ & $\Gamma_{1}$ & $\Gamma_{2}$ & $\Gamma_{3}$   \\
$\Gamma_{3}$ & $\Gamma_{1}$ & $\Gamma_{2}$ & $\Gamma_{3}$   \\
\hline\hline
\end{tabular}
\end{center}
\end{table}

\begin{landscape}
\begin{table}[htb]
\caption{Reductions for the arbitrary case $m$ and $n$ }\label{tableExample}\label{T22}
\begin{center}
\begin{tabular}{c|c|cc}
\hline\hline
Optimal System & Similarity variable && Reductions  \\
\hline
$\Gamma_{1}+c_1\Gamma_{2}+c_2\Gamma_{3}$ & $r=t+c_1x+c_2y, $ & $(XX)$&$c_2 U^{\prime\prime}-c_1c_2\left(n(n-1)U^{n-2}{U^{\prime}}^2+nU^{n-1}U^{\prime\prime}\right)
+
$ \\
&& & ${c_1}^2c_2U^{\prime\prime\prime}+\alpha{c_1}^3V^{\prime\prime\prime}=0$ \\

 & $u=U(r), v=V(r)$ & &$(1-c_1\beta U^m)V^{\prime}-{c_1}^2V^{\prime\prime}=0$ \\
\hline
$\Gamma_{1}$ & $u(t,x,y)=U(x,y),v(t,x,y)=V(x,y)$ & $(XXI)$&$n(n-1)U^{n-2}U_x U_y +n U^{n-1}U_{xy}-U_{xxy}+\alpha V_{xxx}=0$
 \\

& & &$\beta U^m V_x+V_{xx}=0$   \\
\hline
$\Gamma_{2}$ & $u(t,x,y)=U(t,y),v(t,x,y)=V(t,y)$ & $(XXII)$&$U_{ty}=0, $ \\
& & &$V_t=0 $  \\ \hline
$\Gamma_{3}$ & $u(t,x,y)=U(t,x), v(t,x,y)=V(t,x)$ & $(XXIII)$&$\beta U^{m} V_x+V_{xx}-V_t=0
$  \\
 & & &$V_{xxx}=0$  \\ \hline\hline
\end{tabular}
\end{center}
\end{table}
\end{landscape}



\textbf{Reductions and solutions for $(XX)$}\\
\begin{eqnarray}\label{1.5}
V&=&I_3\int\exp\left[\int\left(\frac{1-c_1\beta U^m}{{c_1}^2}\right)\d r\right]\d r,\\
U^\prime+\frac{1}{{c_1}^2}U&=&\frac{1}{{c_1}}U^n-\frac{I_2+I_1r}
{{c_1}^2c_2}+\frac{c_1I_3\alpha}{c_2}\exp\left[\int\left(\frac{1-c_1\beta U^m}{{c_1}^2}\right)\d r\right],
\end{eqnarray}
where $I_1,I_2$ and $I_3$ are constants of integration.\\

\textbf{Reductions and solutions for $(XXI)$ }\\


The system $(XXI)$ admits the following symmetries
$\Gamma_{11}=\partial_x, \Gamma_{12}=\partial_y$. The corresponding optimal system, similarity variables, reductions and solutionsnare given in table \ref{T23}.
%

\begin{table}[htb]
\caption{Solution for equation $(XXI)$}\label{tableExample}\label{T23}
\begin{center}
\begin{tabular}{c|c|c|c}
\hline\hline
Optimal & Similarity  & Reduction & Solution  \\
system&  variable & & \\
\hline
$\Gamma_{11}$ & $U(x,y)=G(y), $ & & $U(x,y)=G(y)$  \\
& $ V(x,y)=H(y)$ & & $V(x,y)=H(y)$   \\
\hline

$\Gamma_{12}$  & $U(x,y)=G(x),$ & $H^{\prime\prime\prime}=0, $&$H=I_1+I_2x+I_3x^2$ \\

 & $V(x,y)=H(x)$ & $\beta G^m H^\prime+H^{\prime\prime}=0$ &$G=\left[-\frac{2I_3}{\beta(I_2+2I_3x)}\right]^{\frac{1}{m}}$ \\
 \hline
$\Gamma_{11}+c_1 \Gamma_{12}$& $r=y-c_1x$, & $n(n-1)G^{n-2}{G^\prime}^2+nG^{n-1}G^{\prime\prime}
+$& $H=I_1\int\exp\left[\frac{\beta}{c_1}\int( G^m )dr\right]dr+I_2$  \\
& & $
+{c_1}G^{\prime\prime\prime}+{c_1}^2\alpha H^{\prime\prime\prime}=0$& $c_1G^{\prime}+G^n=I_3r+I_4-   $  \\
& $U(x,y)=\frac{G(r)}{x}$, & $(\beta G )H^\prime-c_1H^{\prime\prime}=0 $& ${c_1}^2I_1\alpha\exp\left[\frac{\beta}{c_1}\int( G^m) dr\right]$  \\
& $ V(x,y)=H(r)$ & & \\
 \hline\hline

\end{tabular}
\end{center}
\end{table}

%

%

\textbf{Reductions and solutions for $(XXII)$ }\\
The solution of the system $(XXII)$ is given by
\begin{eqnarray}
U(t,y)&=&\int f(t)dt+g(y),\\
V(t,y)&=&h(y).
\end{eqnarray}

\textbf{Reductions and solutions for $(XXIII)$ }\\

The solution of the system $(XXIII)$ is given by
\begin{eqnarray}
U(t,x)&=&\left[\frac{f^\prime(t)+g^\prime(t)x
+h^\prime(t)x^2-2h(t)}{\beta(g(t)+2xh(t))}\right]^{\frac{1}{m}},\\
V(t,y)&=&f(t)+g(t)x+h(t)x^2.
\end{eqnarray}

\section {Conclusion}
In this work, we studied the generalized BLP system by using the theory of Lie symmetries. Specifically, we performed a detailed classification of the admitted Lie point symmetries for the generalized BLP system by constraint the free parameters of the system with the Lie conditions.

We found four different sets for the unknown parameters in which the resulting systems admits different Lie point symmetries. For each system we determined the commutators and the Adjoint representation for the admitted Lie point symmetries. The latter applied to determine the one-dimensional optimal system, an important information in order to determine all the possible and independent similarity transformations which lead to invariant solutions. Finally invariant solutions have been presented for all the cases of study.

This work contributes on the application of Lie's theory on nonlinear differential equations. In a future work we want to investigate the existence of conservation laws which follow from the admitted group properties for the generalized BLP system.

\subsection*{Acknowledgements}
 KK thank Prof. Dr. Stylianos Dimas, S\'ao Jos\'e dos Campos/SP, Brasil for providing new version of SYM-Package.

\end{document}